\documentclass[aps,showpacs,preprintnumbers,amsmath,amssymb]
{revtex4}

\usepackage{graphicx}
\usepackage{dcolumn}
\usepackage{bm}
\usepackage{endnotes}
\begin{document}
\title{Phase transitions of hadronic to quark matter at finite T and $\mu_B$}

\author{B. Liu$^{1,2,3}$,
        M. Di Toro$^{4,5}$ \footnote[1]{ditoro@lns.infn.it},
        G.Y. Shao$^{4}$,
        V. Greco$^{4,5}$, 
        C.W. Shen$^{6}$,    and  Z.H. Li$^{7}$}
\affiliation{
$^{1}$ Center of Theoretical Nuclear Physics, National Laboratory of
 Heavy Ion Accelerator,
       Lanzhou 730000, People's Republic of China\\
$^{2}$ Theoretical Physics Center for Scientific Facilities,
       Chinese Academy of Sciences, Beijing 100049, People's Republic of
China\\
$^{3}$ Institute of High Energy Physics, Chinese Academy of Sciences,
       Beijing 100049, People's Republic of China \\
$^{4}$ Laboratori Nazionali del Sud INFN, I-95123 Catania, Italy\\
$^{5}$ Physcs and Astronomy Dept., University of Catania,  Italy\\
$^{6}$ School of Science, Huzhou Teachers College, Huzhou 313000,
       People's Republic of China\\
$^{7}$ Institute of modern physics, Fudan University, Shanghai 200433,
People's Republic of China}


\begin{abstract}
The phase transition of hadronic to quark matter
and the boundaries of the mixed hadron-quark coexistence phase are studied
within the two Equation of State (EoS) model.
The relativistic effective mean field approach with
constant and density dependent meson-nucleon couplings
is used to describe hadronic matter, and the MIT Bag model is adopted
to describe quark matter. The boundaries of the mixed phase for different
Bag constants are obtained solving the Gibbs equations.
 
We notice that the dependence
on the Bag parameter of the critical temperatures (at zero chemical potential)
can be well
reproduced by a fermion ultrarelativistic quark gas model, 
without contribution from the hadron part.
At variance the critical chemical potentials (at zero temperature) are very
sensitive to the EoS of the hadron sector.
Hence the study of the hadronic EoS is much more relevant for
the determination of the transition to the quark-gluon-plasma at
finite baryon density and low-T. 
Moreover in the low temperature and finite chemical potential region no 
solutions of the Gibbs conditions
are existing for small Bag constant values, $B~<~(135~MeV)^4$. 

 Isospin effects
in asymmetric matter appear relevant in the high chemical potential regions
at lower temperatures, of interest for the inner core properties of
neutron stars and for heavy ion collisions at intermediate energies.
\end{abstract}


\maketitle

\newpage


\section*{1. Introduction}

It has been suggested that a phase transition would take
place from hadronic matter to deconfined quark-gluon matter
at sufficiently high density and/or high temperature.
There has been considerable interest in relativistic heavy-ion collisions
which could offer the possibility of producing a hot and dense matter and/or
plasma of deconfined quarks and gluons.
Up to now, experimental data on the phase transition have been extracted from
ultrarelativistic collisions of almost isospin-symmetric nuclei,
 having a proton fraction Z/A $\sim$ 0.4 - 0.5, in a region of very small
chemical potentials.
 These experimental data from ultrarelativistic collisions
could provide the possibility of studying the phase transition
and testing effective QCD models.

The interest in the transition at high baryon and isospin densities
has been recently growing from the possibility of using new heavy ion
facilities at intermediate energies and for the interest in the properties
of the inner core of neutron stars where the transition to deconfined quark
matter can likely occur.
Now  the main problem is that we do not have
fully reliable effective QCD thoeries able to describe the two phases
so the main approach has been based of a two-EoS model with the Gibbs
conditions. In fact this scheme has been widely used to make
predictions on the phase transition
in the interior of neutron stars~(e.g., se the recent \cite{Shao10,Xu10}
and refs. therein).
We remark that, even in the two-EoS approach, only a few papers have studied
the phase diagram of
hadron-quark transitions at high baryon density in connection to the
phenomenology of
heavy-ion collision in the ten A GeV range (intermediate energies)
\cite{muller,ditoro1,ditoro2,Pagliara10,Cavagnoli10}.

In Ref.~\cite{muller} the phase transition
from hadron to quark matter has been firstly analyzed also for
isospin asymmetric matter.
In this work, a Relativistic Mean Field (RMF)
model, involving the interaction of baryons with isoscalar scalar and vector
fields and
with the isovector $\rho$ meson and pion field, was used for hadronic matter
and a MIT-Bag model involving massless u and d quarks was adopted for
quark matter.
More recently, Refs.\cite{ditoro1, ditoro2}, the RMF
approach was extended to the isovector-scalar $\delta$-field to study
symmetry energy effects on
the possible formation of a mixed hadron-quark phase at high baryon density
 during
intermediate
energy collisions between neutron-rich heavy ions.

We remind that the nonlinear Walecka model ($NLWM$) \cite{SW85,bog97}
based on the $RMF$ effective theory, has been
extensively applied
to study the properties of nuclear matter and neutron star, stable
nuclei and then extended to the drip-line regions.
In the last years some  authors \cite{liubo02,menpro04,baranPR,gait04,liubo05}
have stressed the importance of including the $\delta (a_{0}(980))$ field in
hadronic effective
field theories for asymmetric nuclear matter. The role of the $\delta$ meson
in isospin channels appears relevant at high density regions
\cite{liubo02,menpro04,baranPR,gait04,liubo05}
and so of great interest in nuclear astrophysics.

In order to describe the medium dependence of nuclear interactions,
a density dependent relativistic hadron ($DDRH$) field theory
has been also suggested \cite{fuchs95,jong98,TW99}.
The density dependent meson-nucleon couplings are based
on microscopic Dirac-Brueckner ($DB$) calculations \cite{jong98,vandal04}  and
adjusted to reproduce some nuclear matter and finite nuclei properties
\cite{fuchs95,jong98,TW99}. Recently  the density dependent coupling models
have been  applied to the neutron stars \cite{liubo06}.

In this paper, we use the constant coupling (NL-RMF) scheme and the DDRH model
for hadronic matter and the MIT bag model for quark matter. We study the phase
transition of hadronic to quark matter
and in particular the dependence of the
boundaries of the mixed hadron-quark coexistence phase
 on the choice of the Bag pressure.
Our study leads to the appearance of a Critical-End-Point of the mixed phase
for low values of the Bag parameter. Moreover recently a 
similar effect due to a
density variation of an effective Bag constant has been observed using a
Nambu-Jona Lasinio (NJL) model, with dynamical varying masses for the
quark phase \cite{shao11}.

We also want to see the effects of the different hadronic models on boundaries
of the mixed phase. We find that the critical temperature at low chemical
potential is almost not
affected by variation of the hadron EoS and the Bag-constant dependence
can be well reproduced by a naive ultrarelativistic massless quark gas model.
Conversely the critical chemical potential at vanishing temperature
is very sensitive to the EoS of the hadronic part. Furthermore the mixed phase 
structure at high baryon density is affected by the hadron EoS, especially
in isospin asymmetric matter.
 This has been observed when the
isovector-scalar $\delta$ meson is included and when some density
dependence of the isovector meson couplings are introduced.

The present paper is organized as follows. In order to give more 
emphasis on the new obtained results we report in the Appendices all the 
details about the effective
Equation of State adopted in the two sectors: in Appendix A both models 
used for
hadronic matter
are described while in Appendix B we introduce the MIT-Bag model at finite
temperature for quark matter.

The Gibbs phase transition conditions are presented in Sect.2.
The results are shown and discussed in Sect.3.
Sect.4 is devoted to a detailed analysis of the Bag constant
dependence of critical end points in temperature and chemical
potential.
 Finally some conclusions are drawn in Sect.5.

\section*{2. The phase transition}

According to the Gibbs conditions for phase transition, the temperatures,
the chemical potentials, and pressures of hadronic matter have to be
identical to
that of quark matter inside the mixed phase. Moreover we must require
the conservation of the total baryon and isospin densities :

\noindent
\begin{eqnarray}\label{eq:36}
\mu_{B}^{H}(T, \rho_B^{H}, \rho_3^{H})=\mu_{B}^{Q}(T, \rho_B^{Q}, \rho_3^{Q})~,
\end{eqnarray}

\noindent
\begin{eqnarray}\label{eq:37}
\mu_{3}^{H}(T, \rho_B^{H}, \rho_3^{H})=\mu_{3}^{Q}(T, \rho_B^{Q}, \rho_3^{Q})~,
\end{eqnarray}

\noindent
\begin{eqnarray}\label{eq:38}
P^{H}(T, \rho_B^{H}, \rho_3^{H})=P^{Q}(T, \rho_B^{Q}, \rho_3^{Q})~,
\end{eqnarray}

\noindent
\begin{eqnarray}\label{eq:39}
\rho_{B}^{T}=(1-\chi)\rho_B^H + \chi \rho_B^Q ~,
\end{eqnarray}

\noindent
\begin{eqnarray}\label{eq:40}
\rho_{3}^{T}=(1-\chi)\rho_3^H + \chi \rho_3^Q ~,
\end{eqnarray}

\noindent
 where $\chi$ is the fraction of quark matter in the mixed phase.

The densities and chemical potentials for hadronic matter are defined as

\noindent
\begin{eqnarray}\label{eq:41}
\rho_{B}^{H}=\rho_p + \rho_n~,~~~~
\rho_{3}^{H}=\rho_p - \rho_n~,\\
\mu_{B}^{H}=\frac{1}{2}(\mu_p+\mu_n)~,~~~~\mu_{3}^{H}=\frac{1}{2}
(\mu_p-\mu_n)~,
\end{eqnarray}

and consistently for quark matter we have

\noindent
\begin{eqnarray}\label{eq:42}
\rho_{B}^{Q}=\frac{1}{3}(\rho_u + \rho_d)~,~~~~
\rho_{3}^{Q}=\rho_u - \rho_d~,\\
\mu_{B}^{Q}=\frac{3}{2}(\mu_u+\mu_d)~,~~~~\mu_{3}^{Q}=\frac{1}{2}
(\mu_u-\mu_d)~,
\end{eqnarray}

The asymmetry parameters for hadronic and quark matter are defined,
respectively,

\noindent
\begin{eqnarray}\label{eq:43}
\alpha^{H}=\frac{\rho_n - \rho_p}{\rho_n+\rho_p}~,~~~~
\alpha^{Q}=3\frac{\rho_d - \rho_u}{\rho_d+\rho_u},
\end{eqnarray}

and the total asymmetry is

\noindent
\begin{eqnarray}\label{eq:44}
\alpha^{T}=-\rho_3^{T}/\rho_B^{T}.
\end{eqnarray}

Nucleon and quark chemical potentials, as well as the pressures in the
two phases at the fixed asymmetry and a fixed temperature, can be obtained
directly from
respective EoS, as shown in detail in the Appendices A and B..

In the hadron sector we will use the Non-Linear
Relativistic Mean Field models, \cite{liubo02,baranPR,erice08}, with
different structure of the isovector part, already tested to describe
the isospin dependence of
collective flows and meson production for heavy ion collisions 
at intermediate energies, \cite{greco03,gait04,ferini06}.
We will refer to these different Iso-Lagrangians as: i) $NL$, where no 
isovector
meson is included and the symmetry term is only given by the kinetic 
Fermi contribution, ii) $NL\rho$ when the interaction contribution of
an isovector-vector meson is considered and finally iii) $NL\rho\delta$
where also the contribution of an isovector-scalar meson is accounted for.
 See details in Appendix A1.

We are well aware that there are several uncertainties on the stiffness
of the symmetry energy at high baryon density, mainly due to the lack 
of suitable data, see the reviews \cite{baranPR,baoPR}. Therefore we will 
show also results with
effective hadron interactions based on RMF models with density
dependent meson-nucleon couplings ($DDRH$ forces, Appendix A2)
that present much softer symmetry terms at high baryon density.

As already mentioned, in the quark phase we use the MIT-Bag Model, 
 Appendix B, where 
the symmetry term is only given by the Fermi contribution. An important part
of this work is the study of the Bag-constant dependence of the
transition.


\begin{figure}[hbtp]
\begin{center}
\vglue +3cm
\includegraphics[scale=0.4]{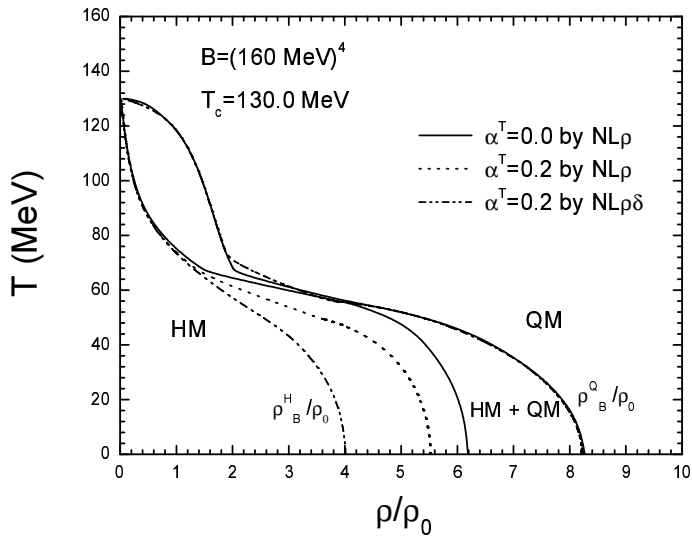}
\vglue -3.0cm
\caption{Phase diagram of hadronic-quark matter coexistence for
symmetric matter ($\alpha^T$=0.0)
and asymmetric matter ($\alpha^T$=0.2) with $B^{1/4}$=160 MeV
in the $NL-RMF$  model as a function of the baryon density.}
\label{trhon160}
\end{center}
\end{figure}

\begin{figure}[hbtp]
\begin{center}
\vglue +3.0cm
\includegraphics[scale=0.4]{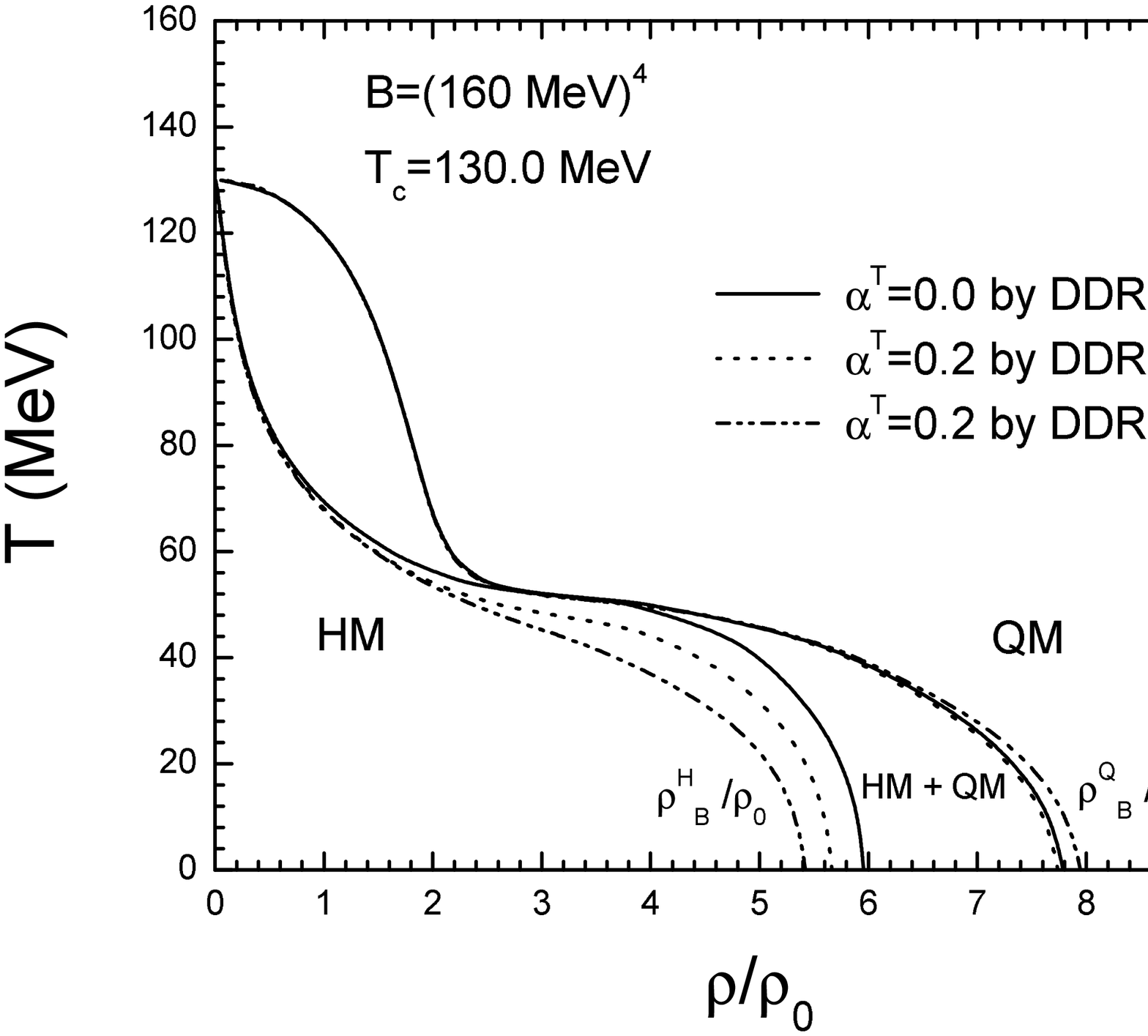}
\vglue -3.0cm
\caption{Same as in Fig.\ref{trhon160} in the $DDRH-RMF$ model.}
\label{trhod160}
\end{center}
\end{figure}

\section*{3. Results: Properties of the Coexistence Zone}



The EOS for hadronic and quark matter is used to calculate the baryon
and the quark chemical potentials, as well as the pressures in the
two phases at the fixed isospin asymmetry and temperature.
We study the phase coexistence region in the (T, $\rho_B$, $\rho_3$) space
and even the properties inside the mixed phase \cite{ditoro2}.
For a fixed value of the total
asymmetry $\alpha^{T}$ and temperature T, we find solutions of 
the $Gibbs$ conditions for both cases of $\chi$=0.0 and 1.0.
In this way the boundary of the mixed phase region
 in the (T, $\rho_B$) plane can be obtained.
Finally we can get the full phase transition diagram by repeating the
procedure.

In our calculations we have used $m_{u}=m_{d}=5.5~ MeV$.
We will only consider the two-flavor case ($q=u,d$) in the bag model
and the interactions between quarks are neglected as a first step.
In fact in this work we are mainly interested in the study of the
Bag Pressure dependence of the results. In order to obtain the
$binodal$ boundaries and the phase diagram, an iterative minimization
procedure is adopted in the search for the solutions of the  $Gibbs$ 
conditions.
In Fig.\ref{trhon160} we present the phase diagram of hadronic-quark matter
coexistence in (T, $\rho$, $\rho_3$) space
with $B^{1/4}$=160 MeV in the $NL\rho$ and the $NL\rho\delta$ models.
In order to see the effects of different $RMF$ models on the phase diagram,
we also use the $DDRH-RMF$ model to describe the hadronic matter with
the same bag constant
to calculate the mixed phase coexistence diagram, the results are given
in Fig.\ref{trhod160}.

From Figs.\ref{trhon160} and \ref{trhod160} we notice that the main
differences between $NL-RMF$
and $DDRH-RMF$ models are the boundary densities
($\rho_B^{Q}/\rho_0$ and $\rho_B^{H}/\rho_0$)
for the quark and hadronic phases at zero temperature
in the isospin asymmetric case.
Both models give almost the same critical temperature
$T_c$=130 MeV,  which then is not
related to hadronic EoS differences as well as to the isospin asymmetry
of the matter.

\begin{figure}[hbtp]
\begin{center}
\vglue +3.0cm
\includegraphics[scale=0.4]{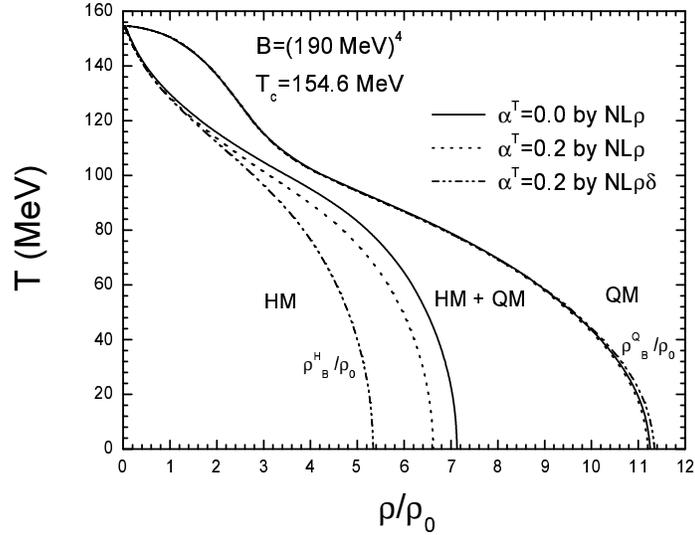}
\vglue -3.0cm
\caption{Same as in Fig.\ref{trhon160} with $B^{1/4}$=190 MeV in
the $NL-RMF$ model.}
\label{trhon190}
\end{center}
\end{figure}



\begin{figure}[hbtp]
\begin{center}
\vglue +3cm
\includegraphics[scale=0.4]{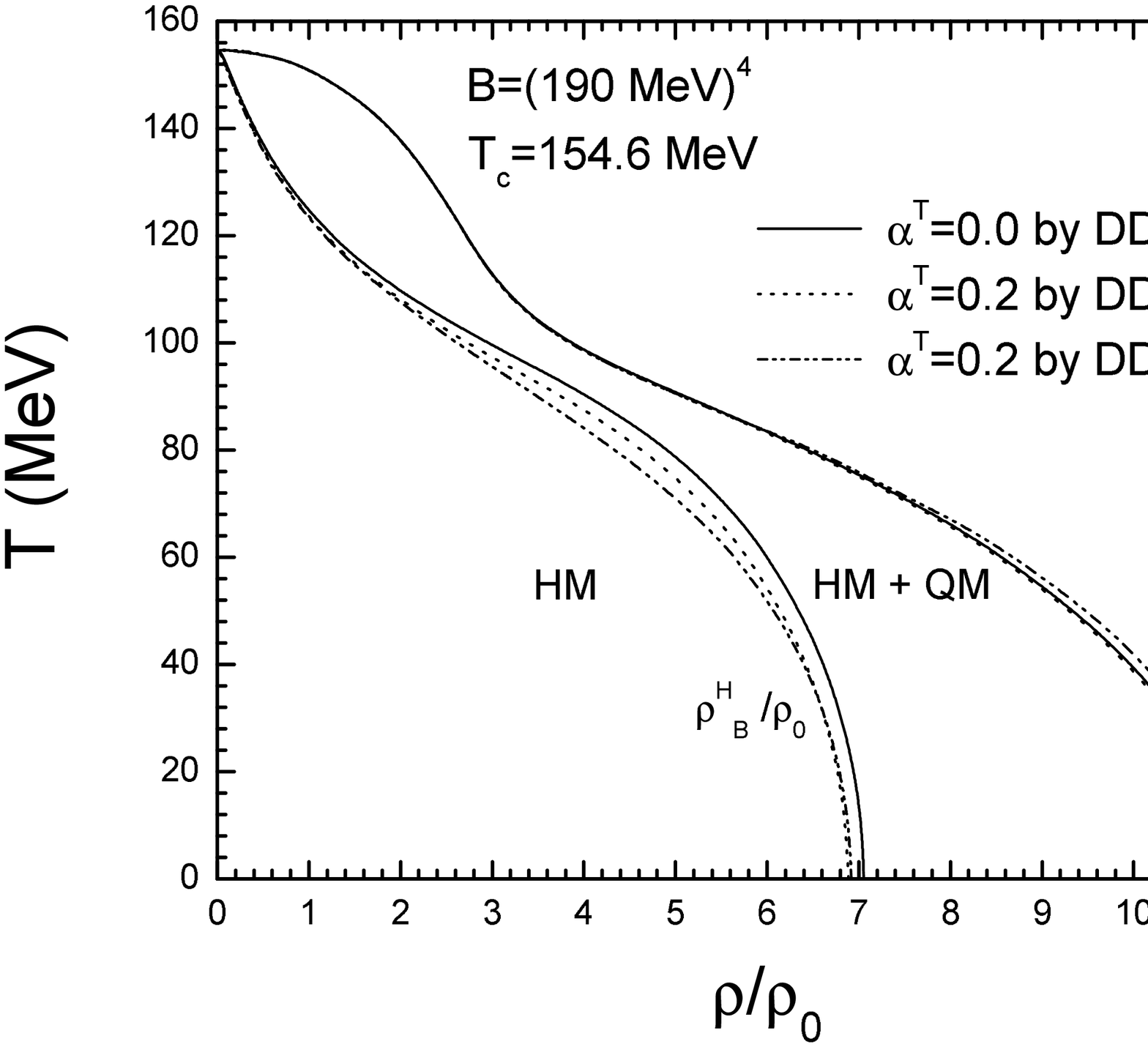}
\vglue -3.0cm
\caption{Same as in Fig.\ref{trhon160} with $B^{1/4}$=190 MeV in the
$DDRH-RMF$
model.}
\label{trhod190}
\end{center}
\end{figure}

In order to know the effects of different values of the bag constant on
the hadronic-quark matter coexistence in (T, $\rho$, $\rho_3$) space,
we use $B^{1/4}$=190 MeV and the two $RMF$ models to calculate the
phase transition diagrams;
the results are shown in Figs.\ref{trhon190} and \ref{trhod190}, respectively.
Like in the previous $B^{1/4}$=160 MeV case,
we can see that the behaviors of the binodal boundaries for symmetric 
matter in the quark
and hadronic phases in (T, $\rho$) space given by the both $NL-RMF$
and $DDRH-RMF$ models are roughly the same.

At variance, for asymmetric matter interesting isospin effects are appearing
on the critical points at high baryon density and low temperatures, 
directly related to
the high density behavior of the symmetry energy in the hadron phase
 \cite{ditoro2,shao11}, see also the discussion in Appendix A.
In fact we get that for both Bag-constant choices
the boundaries of $\alpha^{T}$=0.2 in (T, $\rho$) space for the hadronic
phase given by
$DDRH\rho$ and $DDRH\rho\delta$ models are different from that
given by $NL\rho$ and $NL\rho\delta$ models.

\begin{figure}[hbtp]
\begin{center}
\vglue -2.0cm
\includegraphics[scale=0.7]{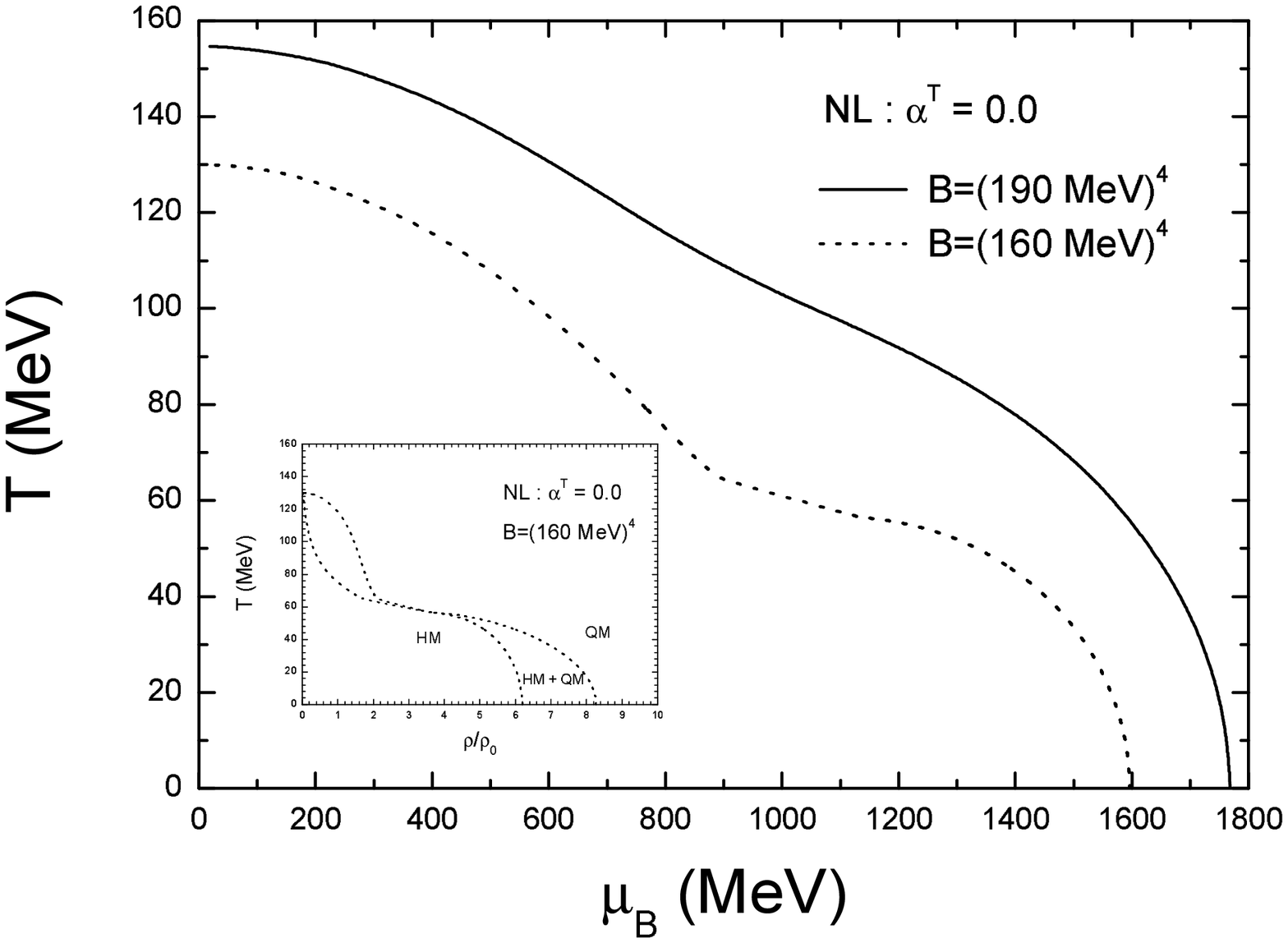}
\vglue -7.0cm
\caption{$T-\mu_B$ phase diagram in the $NL\rho$ model
with $B^{1/4}$=160 and 190 MeV.
The inset is the $T-\rho_B$ phase diagram in the $NL\rho$
model with $B^{1/4}$=160 MeV}
\label{tmunl}
\end{center}
\end{figure}

From Figs. \ref{trhon160}, \ref{trhod160},
 \ref{trhon190} and  \ref{trhod190}, we remark that some
squeezing-reopening effects of the binodal surface appear.
The squeezing-reopening effect is more evident in the $B^{1/4}$= 160 MeV
case and
this is due to the variation of the pressure in the quark phase, for a
fixed chemical potential $\mu_B^{Q}$, as shown in the following.

In order to simplify the discussion we consider the case of symmetric
matter, $\alpha^{T}$=0.0,
and only the hadronic $NL$ interaction.
First we define the transition chemical potential
$\mu_B^{tr}$ $\equiv$ $\mu_B^{H}$=$\mu_B^{Q}$.
In Fig.\ref{tmunl}
we plot the transition phase diagrams for symmetric matter
in the (T, $\mu_B$) space with  $B^{1/4}$= 160 and 190 MeV, in
the $NL$ model.
The inset is the corresponding phase diagram of the transition  in the 
(T, $\rho$) space
 for symmetric matter with $B^{1/4}$= 160 MeV.
In order to discuss the squeezing-reopening effect both (T, $\mu_B$)
and (T, $\rho$) phase diagrams should be considered.

 If we look at the (T, $\rho$)
inset in Fig.\ref{tmunl}, we note that
 three main regions exist :\\

(1) The ``opening'' region at low baryon densities and high temperatures.
We can see from Fig.\ref{tmunl} that the transition chemical potential
$\mu_B^{tr}$ is smaller than the
effective nucleon mass and the effective chemical potentials
$\mu_i^{\star}$ even smaller in this region.
This indicates that $E_i^{\star}$ $>$ $\mu_i^{\star}$ for the hadronic
part and the nucleon fermion distribution
will be small. So the $\rho_B^{H}$ will present a slow increase with
$\mu_B$ ($f_{i}(k)$ $\simeq$ $\bar{f}_{i}(k))$. We note that the hadron
pressure is mostly of thermal nature and
with a positive antifermion contribution in this region.
The $\mu_B$ in the quark part is always much larger than the current quark
masses ($m_u=m_d$=5.5 MeV)
and so we can have a fast increase of the $\rho_B^{Q}$.

(2) The ``re-opening'' region at high baryon densities and low temperatures.
In this region, $\mu_B$ is larger than the nucleon effective mass.
 This means that $E_i^{\star}$ $<$ $\mu_i^{\star}$ for the hadronic part,
the nucleon fermion distribution will not be small and so the $\rho_B^{H}$ will
show a fast increase with $\mu_B$, also because the antifermion negative
contribution is of course
reduced. The interaction part of the hadron pressure shall increase.
In correspondence we need a fast increase also of the $\rho_B^{Q}$ in order
to keep the $P^{H}=P^{Q}$ balance.

(3) The plateau at intermediate region of baryon densities and temperatures.
In this region we get $\mu_B$ $\simeq$ $M^{\star}$ (nucleon effective mass),
thus we can have the squeezing of the mixed phase in the $T -\rho_B$ plane and
a kind of plateau in the $T-\mu_B$ plane.
Our study clearly shows that the squeezing-reopening effect on the mixed phase
is very sensitive to the value of the MIT bag constant.

\begin{figure}[hbtp]
\begin{center}
\vglue +1.0cm
\includegraphics[scale=0.35]{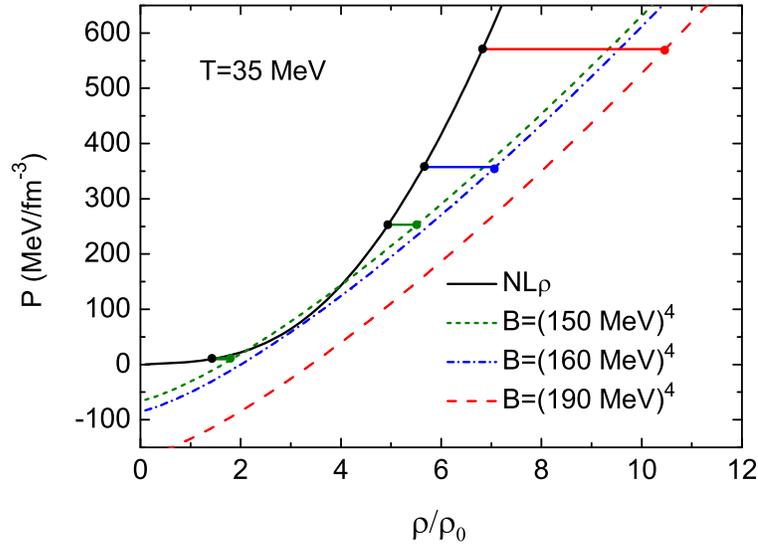}
\caption{$P-\rho_B$ phase diagram of hadronic-quark matter for symmetric matter
($\alpha_T$=0.0) at temperature $T~=~35~MeV$, obtained 
by the $NL\rho$  model and various choices of the MIT-Bag constants.}
\label{prho35}
\end{center}
\end{figure}

Finally in Fig.\ref{prho35}
we present the pressure-baryon density phase diagram of the hadron-quark 
transition for symmetric
 matter
($\alpha_T$=0.0) always in the $NL$  model and for various choices of the
MIT-Bag constant. The temperature is fixed at $35~MeV$, i.e. in the region
of the mixed phase squeezing. The Maxwell constructions are indicated by  
the connections  between the solid line of the hadron pressure and the 
dashed, dot-dashed
and dotted lines of the quark pressure. It can be seen that the 
squeezing-reopening effect
 is larger for lower the bag constant choices.
 In fact for small values of the bag constant
 we cannot have the solution in the intermediate, flattening, region.
This  means that the Gibbs conditions cannot be satisfied and we get 
something like a critical End-Point of the mixed phase. From the figure 
we see that this is first happening for a Bag constant 
$B^{1/4}~\simeq~155~MeV$.
If we further decrease the Bag pressure we reach the limit of no solutions
in general at high chemical potential, since the baryon pressure cannot match
the quark pressure, and the mixed phase just disappears.
This important result at low temperatures and high densities will be further 
discussed in detail in the next Section, in particular in connection to the 
Fig.\ref{PMu-T0}.


\begin{figure}[hbtp]
\begin{center}
\vglue +3.0cm
\includegraphics[scale=0.4]{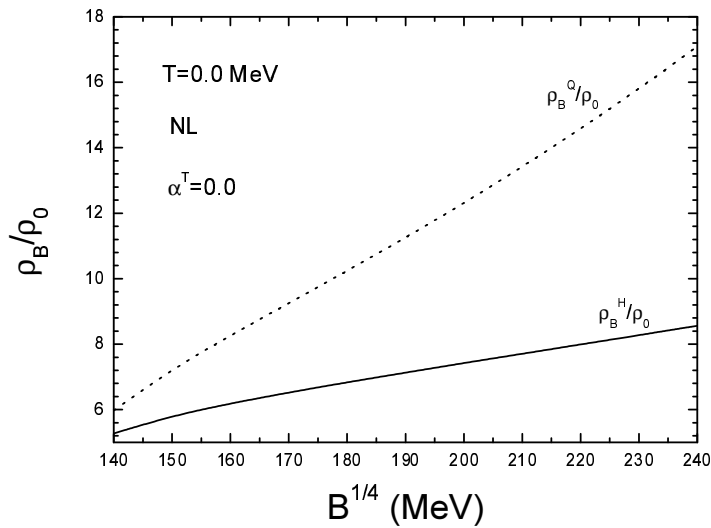}
\vglue -4.0cm
\caption{Baryon and quark densities as a function of the bag
constant at T=0.}
\label{rhoB}
\end{center}
\end{figure}

\section*{4. Dependence of the binodal surface on the Bag constant}

We will focus our analysis on results for isospin symmetric matter.
 In Fig.\ref{rhoB} we plot the density boundaries of the binodal region at
T=0 as a function of
the bag constant $B^{1/4}$ in the interval 140-240 MeV.
The dotted line is the quark matter limit density $\rho_B^{Q}/\rho_0$,
the solid line is the hadronic matter limit density $\rho_B^{H}/\rho_0$.
We clearly see that both limits are increasing with larger Bag values, as
expected from the fact the the energy per particle in the quark phase is
increasing.
It can be also seen that $\rho_B^{Q}/\rho_0$ increases more quickly
with the increasing Bag constant than the $\rho_B^{H}/\rho_0$.
So the selection of the Bag constant value will influence the binodal
surface and in general the final
results of the phase transition.

This discussion about Bag constant effects on the transition is more
clear in the ($T,\mu_B$) plane.
Already from Fig.\ref{tmunl} (for symmetric matter), where results are
reported for $B^{1/4}$= 160 and 190 MeV values, it is evident that
the transition region is deeply affected by the choice of the Bag constant,
in particular at the end points at zero temperature or zero chemical potential.
We will separately analyze the two regions.


\begin{figure}[hbtp]
\begin{center}
\vglue +4.0cm
\includegraphics[scale=0.4]{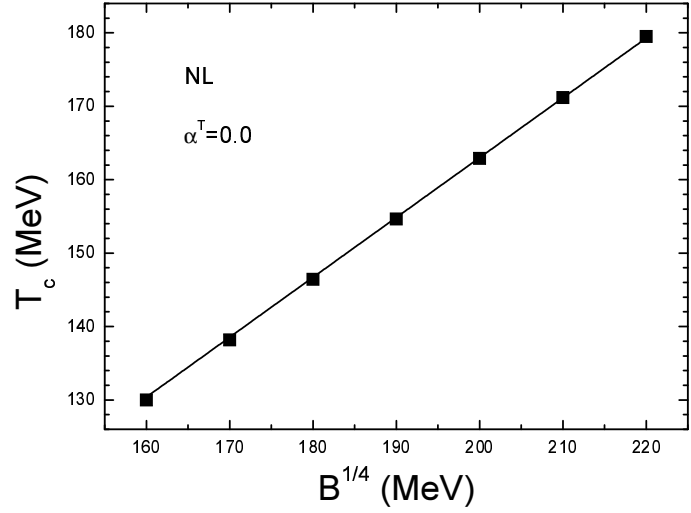}
\vglue -4.0cm
\caption{Critical temperature for symmetric matter ($\alpha_T$=0.0),
at zero chemical potential ($\mu_B=0$),
as a function of the bag constant $B^{1/4}$.
 Black Squares: calculated values from Gibbs conditions. Solid line:
 $T_c=0.815 B^{1/4}$ fit.}
\label{tcB}
\end{center}
\end{figure}


\begin{figure}[hbtp]
\begin{center}
\vglue +3.0cm
\includegraphics[scale=0.4]{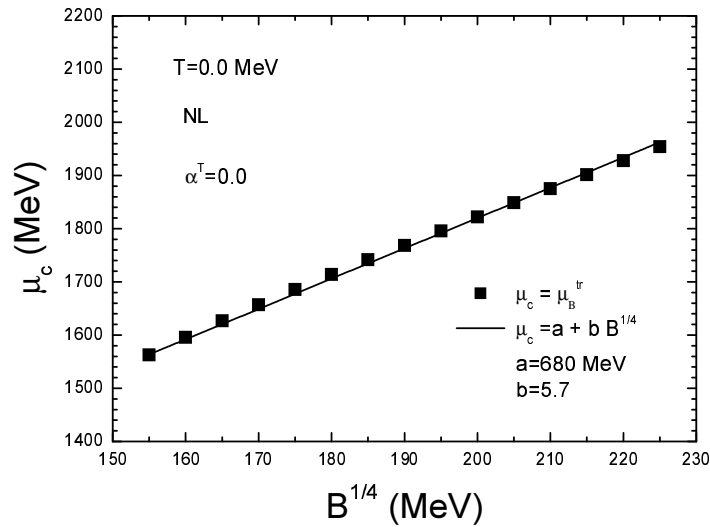}
\vglue -4.0cm
\caption{Critical chemical potential for symmetric matter ($\alpha_T$=0.0),
at zero temperature,
as a function of the bag constant $B^{1/4}$.}
\label{mubag}
\end{center}
\end{figure}

In Fig.\ref{tcB} we report the critical temperatures for symmetric matter,
at zero chemical potential,
for different values of the MIT Bag constant in the 160-220 MeV interval,
using the $NL$ model
for the hadronic sector.
The square points are the calculated values while
the solid line is given by the simple linear fit : $T_c=0.815 B^{1/4}$.

Similarly in the Fig.\ref{mubag} we show the critical chemical
potentials, at zero temperature,
for different values of the MIT Bag constant in the same 160-220 MeV interval,
always for symmetric matter
in the $NL$ model
for the hadronic sector.
The square points are the calculated values while
the solid line corresponds, even in this case, to a rather simple linear
fit : $\mu_c= a + b B^{1/4}$, with $a = 680 MeV$ and $b = 5.7$.
It is interesting to analyze such almost linear behavior of the two end-points
as a function of the Bag values, see the following subsection.

Finally, as already noted,
we must remark that no $\mu_c~\not=~0$ solutions
 are existing at zero temperature
 for very small Bag constants, $B~<~(135~MeV)^4$. This is due to the fact
that at low temperature and final chemical potential the hadron pressure,
cannot compensate the quark pressure in the coexistence region,
 as we can also see from Fig.\ref{PMu-T0}.
Of course this effect is depending on the EoS in the hadron sector. This point 
will also be discussed in the next subsection.

\subsection*{4.1 Ultrarelativistic gas model}

For the quark sector we can use as a reference the naive fermion 
ultrarelativistic gas model,
considering only free quarks in the Bag. Consistently with the results
obtained before, no gluons are included.
The pressure is simply given by \cite{Bmuller85,Yagi05}

\begin{equation}\label{pquark}
P^Q = \frac{(g_Q+g_{\bar Q})}{3} T^4 [ \frac{7 \pi^2}{120} + \frac{1}{4} 
(\frac{\mu}{T})^2
 + \frac{1}{8 \pi^2}(\frac{\mu}{T})^4 ] - B
\end{equation}

where $g_Q~=~g_{\bar Q}~=~12$ is the spin-color-flavor degeneracy factor.
The condition $P^Q~=~P^H$ will give the critical line of the transition
to the quark deconfined matter. 
Neglecting the hadron contribution ($P^Q~=~0$ condition), from the 
Eq.(\ref{pquark}) we can easily get the two critical end points,
at zero chemical potential and at zero temperature, as a function of the
Bag parameter.

At $\mu = 0$ we get a critical temperature

\begin{equation}\label{tcrit}
T_c = [\frac{30}{7 \pi^2}]^{1/4} B^{1/4} = 0.81 B^{1/4}~MeV
\end{equation}

 in a very good agreement with
the results obtained from the Gibbs conditions, solid line in Fig.\ref{tcB}.
The hadron EoS is not contributing to the critical temperature at zero chemical
potential, as we can also clearly see from the Figs.\ref{trhon160}
and \ref{trhod160}. The reason is that at $\mu=0$ only thermal vacuum 
excitations contribute to the pressure, ruled essentially by the particle 
degrees of freedom, which are much larger for the quark phase \cite{Yagi05}.

\begin{figure}[hbtp]
\begin{center}
\includegraphics[scale=0.4]{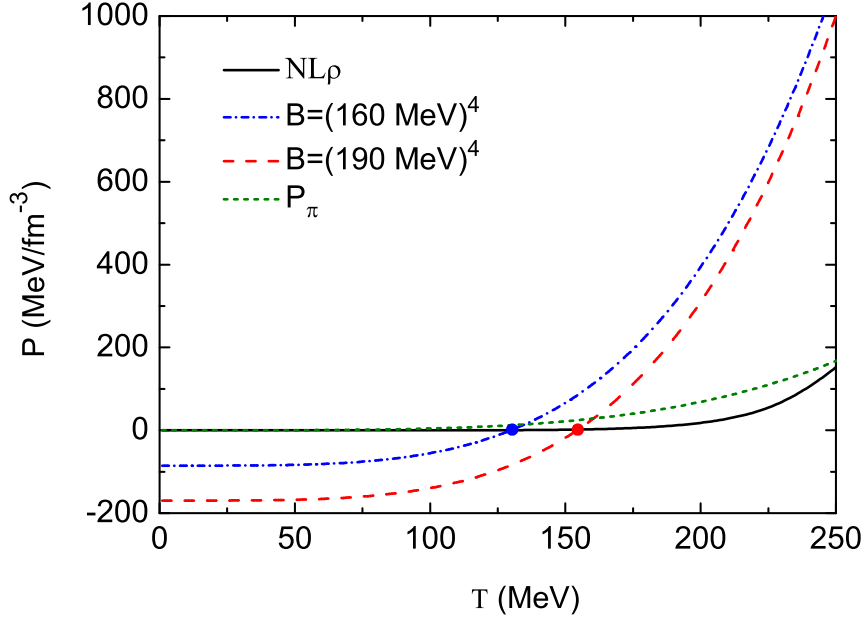}
\caption{Symmetric matter ($\alpha_T$=0.0)
at zero chemical potential: Hadron pressure (solid, NL choice) and Quark 
pressure $P^Q$, Eq.(\ref{pquark}), for various Bag constants 
(dashed, dot-dashed) 
as a function of temperature. The dotted line gives the pressure of
a massless pion gas, see text.
The crossing points correspond to the critical 
temperatures.}
\label{PT-mu0}
\end{center}
\end{figure}

This is nicely confirmed by the Fig.\ref{PT-mu0} where we plot the 
temperature evolution of the pressure in the two phases at $\mu~=~0$. We see 
that the hadron pressure is almost negligible up to the crossing points
(the $T_c$ obtained from the full Gibbs conditions). 
We can estimate an 
increase at much larger temperatures, due to the onset of anti-hadron 
contributions, also related to a decrease of the effective nucleon masses.. 

Of course at $\mu = 0$ one can argue that the nucleon 
mass is too high to be thermally excited and in fact it is known that the
hadronic matter is now dominated by pions. 
It is then interesting to compare the hadron pressure obtained 
from the $RMF$ approach with the results of a pure massless pion gas 
pressure $P_\pi$, dotted line of Fig.\ref{PT-mu0}, given by \cite{Yagi05}
$$
P_\pi = g_\pi \frac{\pi^2}{90} T^4
$$
where $g_\pi = 3$ is the pion degeneracy. The critical temperatures are almost
not affected. A more refined model would be to include all the mesonic 
resonances with their masses. However this does not significantly change 
the hadronic pressure up to $T~\simeq~150-160~MeV$ \cite{venu92,bora10}.
 We nicely see that the pure thermal hadron pressure 
is only ruled by the particle degrees of freedom, as already noted.

\vskip 0.2cm

At variance if we perform the same analysis at
$T = 0$ in order to get the B-dependence of the critical quark chemical 
potentials we get

\begin{equation}\label{mucrit}
\mu_c = [2 \pi^2]^{1/4} B^{1/4} = 2.11 B^{1/4} ~ MeV
\end{equation}

which corresponds to baryon chemical potential, Eq.(\ref{eq:35}),

\begin{equation}\label{mubcrit}
\mu_c^B = \frac{3}{2} (\mu_c^u + \mu_c^d) = 3 \mu_c = 6.33 B^{1/4} ~ MeV
\end{equation}

which is in large disagreement with
the results obtained from the Gibbs conditions, solid line
of Fig.\ref{mubag}, corresponding to a
linear
fit $\mu_c= a + b B^{1/4}$, with $a = 680~MeV$ and $b = 5.7$,
 we note two differences: i) The  $ B^{1/4}$ slope is below the
$6.33$ value expected by the pure quark gas model; ii) We need a constant term
$a = 680~MeV$. Both points are related to density dependence of the hadron
interaction, that now cannot be neglected. The different slope comes from 
interaction contributions
to the pressure
in the hadron phase. The added constant is due to the presence of the nucleon
rest mass in the hadron chemical potential. It is difficult to work out an
analytical derivation of these effects since the relation between pressure 
and chemical potential in the hadron sector is not simple and naive Fermi gas 
models are largely overestimating the pressures in $P~-~\mu_B$ plots
\cite{pressure}.
We can only state that at 
zero temperature and large chemical potentials the critical points are 
very sensitive to the hadron contribution.

\begin{figure}[hbtp]
\begin{center}
\includegraphics[scale=0.4]{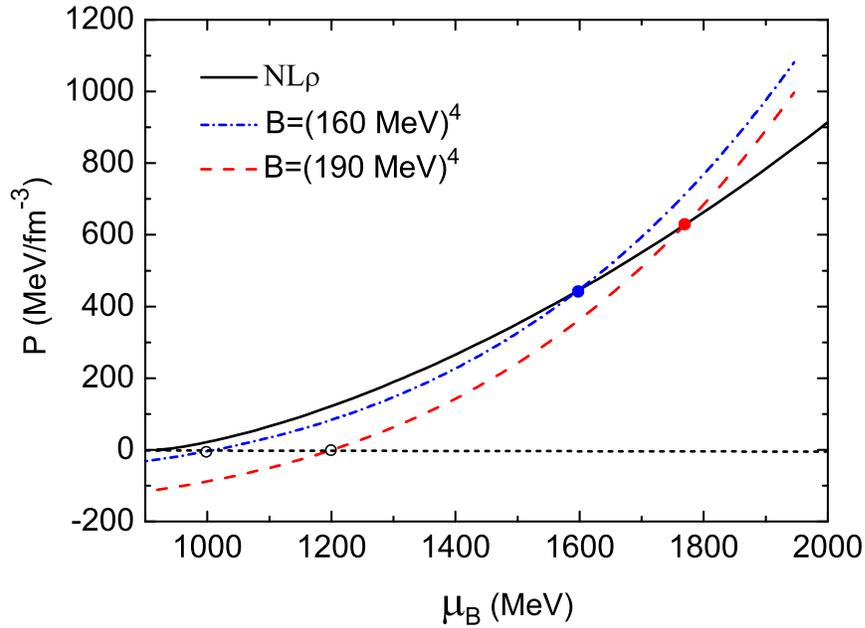}
\caption{Symmetric matter ($\alpha_T$=0.0)
at zero temperature: Hadron pressure (solid, NL choice) and Quark 
pressure $P^Q$ for various Bag constants (dashed, dot-dashed) 
as a function of baryon chemical potential. We show also $P^Q~=~0$
line corresponding to the case without hadron contribution. 
The crossing points (solid with the hadron part, open without the hadron
contribution) correspond to the critical 
chemical potentials.}
\label{PMu-T0}
\end{center}
\end{figure}

This can be easily seen from Fig.\ref{PMu-T0} where we show the fully 
calculated pressures $P^Q$ and $P^H$ {\it vs.} the baryon chemical potential 
at zero temperature. We show also the points on the line $P^Q~=~0$ 
corresponding to the 
deconfinement conditions without hadron contributions. We clearly notice the 
difference 
in the crossing points with and without the hadron part. The critical chemical
potentials get rather larger values (of about a $50\%$) with the hadron 
contribution and also the 
B-dependence looks different.

Finally we can work out a simple consistency check of the results shown in
the Figs. \ref{rhoB} and \ref{mubag} using the zero temperature relationship
between quark chemical potentials and densities $\rho_q~=~\mu_q^3/\pi^2$.
Since $\rho_B^Q~=~\frac{1}{3}(\rho_d+\rho_u)~=~\frac{2}{3} \rho_q$ we finaly
get
$$
\mu_B^Q = 3 \mu_q = 3 (\frac{3 \pi^2 \rho_0}{2})^{1/3}
(\frac{\rho_B^Q}{\rho_0})^{1/3} (\hbar c) ~ MeV
$$

Using the critical quark chemical potential parametrization of
Fig.\ref{mubag}, $\mu_c= a + b B^{1/4}$, with $a = 680~ MeV$ and $b = 5.7$,
we nicely get the $B^{1/4}$ dependence of the quark baryon density limit
line of the coexistence region shown in Fig.\ref{rhoB}.

\section*{5. Concluding remarks}

We use the two kinds of RMF models for hadronic matter, with constant and
density dependent couplings,
and the MIT-Bag model for the quark matter
to study the boundaries of the mixed coexistence phase
in the transition from hadronic to quark matter.
Binodal surfaces and critical end-points for different isospin asymmetries
are obtained using different hadronic models and various MIT Bag constants.
 The Bag Pressure dependence of the critical points is very interesting since
we clearly show the difference between two regions:
\begin{itemize}
\item{High Temperature and Low Baryon Chemical Potential.

The hadron EoS is not important: we get the same $T_c$ for any hadron EoS, 
even including a pure massless pion gas contribution. Isospin effects
in asymmetric matter are also negligible. 
In conclusion the critical temperature at zero chemical potential, when 
existing,
only depends on the value of the Bag constant. In fact we show that if small
Bag values are used the binodal surface is shrinking at low densities and
finite temperatures and eventually no continous solutions 
down to  $\mu~=~0$
are found
for  $B~<~(150~MeV)^4$. 

We note that a $T_c$ comparison with recent lattice
QCD calculations with physical quark masses \cite{bora10} would indicate 
an effective Bag
constant, at low $\mu$, around $B~\simeq~(180-190~MeV)^4$.}
\item{High Baryon Chemical Potential and Low Temperature.

Now we have just the opposite evidence: the Critical Chemical Potential
$\mu_c$ is clearly dependent on the interaction in the hadron phase.
Actually we even show that if small
Bag values are used the binodal surface is shrinking at high densities and
finite temperatures and eventually no $\mu~\not=~0$ solutions are found
for  $B~<~(135~MeV)^4$.
 
Moreover for asymmetric matter all that translates into relevant isospin 
effects, 
very sensitive to different
symmetry energy terms in the Hadron EoS. 
These results are of interest for 
all nuclear systems at high baryon and isospin  density and moderate 
temperature, like neutron star formation and inner core structure 
and/or heavy ion collisions at intermediate energies.}

\end{itemize}

\section*{Acknowledgments}

This work has been partially performed under the FIRB Research Grant
RBFR0814TT of the Italian Minister of Education, Universities and Research.
This project is supported by the National Natural Science Foundation of China
under Grant Nos. 10875160, 11075037, 10979024, 10905021,
the Key Project of Science and Technology Research of Ministry of
Education of China
under grant number 209053,
the Natural Science Foundation of Zhejiang Province of China under grant
number Y6090210 and the INFN of Italy.

\vspace{1.5cm}

\appendix

\section{Equation of state for hadronic matter at finite temperature}

\subsection*{A1. Nonlinear relativistic mean field model with constant
couplings}


The Lagrangian density, with the isovector scalar $\delta$  field,
 used in this work is

\noindent
\begin{eqnarray}\label{eq:1}
{\cal L } &=& \bar{\psi}[i\gamma_{\mu}\partial^{\mu}-(M-
g_{\sigma}\phi -g_{\delta}\vec{\tau}\cdot\vec{\delta})
-g{_\omega}\gamma_\mu\omega^{\mu}-g_\rho\gamma^{\mu}\vec\tau\cdot
\vec{b}_{\mu}]\psi \nonumber \\&&
+\frac{1}{2}(\partial_{\mu}\phi\partial^{\mu}\phi-m_{\sigma}^2\phi^2)
-U(\phi)+\frac{1}{2}m^2_{\omega}\omega_{\mu} \omega^{\mu}
+\frac{1}{2}m^2_{\rho}\vec{b}_{\mu}\cdot\vec{b}^{\mu} \nonumber
\\&&
+\frac{1}{2}(\partial_{\mu}\vec{\delta}\cdot\partial^{\mu}\vec{\delta}
-m_{\delta}^2\vec{\delta^2}) -\frac{1}{4}F_{\mu\nu}F^{\mu\nu}
-\frac{1}{4}\vec{G}_{\mu\nu}\vec{G}^{\mu\nu},
\end{eqnarray}

\noindent
where $\phi$ is the $\phi$-meson field,
$\omega_{\mu}$ is the $\omega$-meson field, $\vec{b}_{\mu}$ is
$\rho$ meson field, $\vec{\delta}$ is the isovector scalar field of the
$\delta$-meson.
$F_{\mu\nu}\equiv\partial_{\mu}\omega_{\nu}-\partial_{\nu}\omega_{\mu}$,
$\vec{G}_{\mu\nu}\equiv\partial_{\mu}\vec{b}_{\nu}-
\partial_{\nu}\vec{b}_{\mu}$,
and the $U(\phi)$ is a nonlinear potential of $\sigma$ meson :
$U(\phi)=\frac{1}{3}a\phi^{3}+\frac{1}{4}b\phi^{4}$.

The EOS for nuclear matter with the isovector scalar field
 at finite temperature in the mean-field approximation (MFA) is given by

\noindent
\begin{eqnarray}\label{eq:2}
\epsilon= 2 \sum_{i=n,p}\int \frac{{\rm d}^3k}{(2\pi)^3}E_{i}^\star(k)
(f_{i}(k)+\bar{f}_{i}(k))
+\frac{1}{2}m_\sigma^{2}\phi^2
+ U(\phi) \nonumber \\
+\frac{1}{2}\frac{g_\omega^2}{m_\omega^{2}}\rho^2
+\frac{1}{2}\frac{g_\rho^2}{m_\rho^{2}}\rho_3^2
+\frac{1}{2}\frac{g_\delta^2}{m_\delta^{2}}\rho_{s3}^{2}~,
\end{eqnarray}

\noindent
\begin{eqnarray}\label{eq:3}
 P =\frac{2}{3}\sum_{i=n,p}\int \frac{{\rm d}^3k}{(2\pi)^3}
\frac{k^2}{E_{i}^\star(k)} (f_{i}(k)+\bar{f}_{i}(k))
-\frac{1}{2}m_\sigma^2\phi^2
-U(\phi) \nonumber \\
+\frac{1}{2}\frac{g_\omega^2}{m_\omega^{2}}\rho^2
+\frac{1}{2}\frac{g_\rho^2}{m_\rho^{2}}\rho_3^2
-\frac{1}{2}\frac{g_\delta^2}{m_\delta^{2}}\rho_{s3}^{2}~,
\end{eqnarray}

where ${E_i}^\star=\sqrt{k^2+{{M_i}^\star}^2}$. The nucleon effective masses
are defined as

\noindent
\begin{equation}\label{eq.4}
{M_i}^\star=M-g_\sigma\phi\mp g_\delta\delta_3~~~ (-~proton, +~neutron).
\end{equation}

The field equations in the relativistic mean field (RMF) approach are

\noindent
\begin{equation}\label{eq.5}
\phi=-\frac{a}{m_\sigma^2}\phi^2-\frac{b}{m_\sigma^2}\phi^3
+\frac{g_\sigma}{m_\sigma^2}(\rho_{sp}+\rho_{sn})~,
\end{equation}

\noindent

\begin{equation}\label{eq.6}
\omega_0=\frac{g_\omega}{m_\omega^2}\rho~,
\end{equation}

\noindent
\begin{equation}\label{eq.7}
b_0=\frac{g_\rho}{m_\rho^2}\rho_3~,
\end{equation}

\noindent
\begin{equation}\label{eq.8}
\delta_3=\frac{g_\delta}{m_\delta^2}(\rho_{sp}-\rho_{sn})~,
\end{equation}

\noindent
with the baryon density $\rho\equiv \rho_{B}^{H}=\rho_{p}+\rho_{n}$,
  $\rho_3\equiv \rho_3^H=\rho_p-\rho_n$ and $\rho_{s3}=\rho_{sp}-\rho_{sn}$,
$\rho_{sp}$ and $\rho_{sn}$ are the scalar densities for proton and neutron,
respectively.
The $f_i(k)$ and $\bar{f}_{i}(k)$ in Eqs.(A2)-(A3) are the fermion and
antifermion distribution functions for protons  and neutrons ($i=p,n$):

\noindent
\begin{eqnarray}\label{eq.9}
f_i(k)=\frac{1}{1+\exp\{[{E_i}^\star(k)-{\mu_i^\star}]/T \} }\,,
\end{eqnarray}

and

\begin{eqnarray}\label{eq.10}
\bar{f}_{i}(k)=\frac{1}{1+\exp\{[{E_i}^\star(k)+{\mu_i^\star}]/T \} }.
\end{eqnarray}

\noindent
where the effective chemical potential ${\mu_{i}}^\star$
is determined by the nucleon density $\rho_{i}$

\noindent
\begin{eqnarray}\label{eq:11}
\rho_i=2\int\frac{{\rm d}^3k}{(2\pi)^3}(f_{i}(k)-\bar{f}_{i}(k))\,,
\end{eqnarray}

\noindent
and the $\mu_i^\star$ is related to the chemical potential $\mu_i$ in terms
of the vector meson mean fields by the equation

\noindent
\begin{eqnarray}\label{eq.12}
\mu_i^\star ={\mu_i} - g_\omega\omega_0\mp g_{\rho}b_0~~~(-~proton, +~neutron),
\end{eqnarray}

\noindent
where $\mu_i$ are the thermodynamical chemical potentials
$\mu_i=\partial\epsilon/\partial\rho_i$.  The chemical potentials for proton
and neutron are given by,
respectively

\noindent
\begin{eqnarray}\label{eq.13}
\mu_p=\mu_p^\star + \frac{g_\omega^2}{m_\omega^2}\rho + \frac{g_\rho^2}
{m_\rho^2}\rho_3 \nonumber \\
\mu_n=\mu_n^\star + \frac{g_\omega^2}{m_\omega^2}\rho - \frac{g_\rho^2}
{m_\rho^2}\rho_3.
\end{eqnarray}

The proton and neutron chemical potentials can be denoted in terms of the
baryon and isospin
chemical potentials by the equations

\noindent
\begin{eqnarray}\label{eq.14}
\mu_p =\mu_B + \mu_3, ~~~~~\mu_n =\mu_B - \mu_3.
\end{eqnarray}

\noindent
The scalar density $\rho_{s}$ is given by

\noindent
\begin{eqnarray}\label{eq:15}
\rho_{s}=2\sum_{i=n,p}\int\frac{{\rm d}^3k}{(2\pi)^3}\frac{M_{i}^{\star}}
{E_{i}^{\star}}(f_{i}(k)+ \bar{f}_{i}(k))~.
\end{eqnarray}


\subsubsection*{A1.1 Parameters}

{\hskip  0.7cm}
The isovector coupling constants, $\rho$-field and $\rho+\delta$ cases,
are fixed from the symmetry energy
at saturation and from Dirac-Brueckner estimations, see the detailed
discussions in refs.\cite{liubo02,greco}.

The coupling constants, $f_{i}\equiv g_{i}^{2}/m_{i}^{2}$,
$i=\sigma, \omega, \rho, \delta$, and the two parameters of the $\sigma$
self-interacting terms : $A\equiv a/g_{\sigma}^{3}$ and $B\equiv
b/g_{\sigma}^{4}$ are reported in Table 1.
The corresponding properties of nuclear matter are listed in Table 2.
Here the energy per nucleon is defined $E/A=\epsilon/\rho-M$.

\par
\vspace{0.3cm}
\noindent

\begin{center}
{{\large \bf Table 1.}~Parameter set.}
\par
\vspace{0.5cm}
\noindent

\begin{tabular}{c|c|c} \hline
$Parameter~~Set$  &NL$\rho$  &NL$\rho\delta$ \\ \hline
$f_\sigma~(fm^2)$  &10.329   &10.329    \\ \hline
$f_\omega~(fm^2)$  &5.423    &5.423     \\ \hline
$f_\rho~(fm^2)$    &0.95     &3.150      \\ \hline
$f_\delta~(fm^2)$  &0.00     &2.500       \\ \hline
$A~(fm^{-1})$      &0.033    &0.033       \\ \hline
$B$                &-0.0048  &-0.0048     \\ \hline
\end{tabular}
\end{center}
\vspace{0.5cm}
\noindent


\begin{center}
{{\large \bf Table 2.}~Saturation properties of nuclear matter.}

\par
\vspace{0.5cm}

\noindent
\begin{tabular}{ c c c } \hline
$\rho_{0}~(fm^{-3})$ &0.16   \\ \hline
$E/A ~(MeV)$         &-16.0  \\ \hline
$K~(MeV)$            &240.0   \\ \hline
$E_{sym}~(MeV)$      &31.3    \\ \hline
$M^{*}/M $           &0.75    \\ \hline
\end{tabular}
\end{center}

\subsection*{A2 Density Dependent coupling model}


The  Lagrangian density, with $\delta$ meson, now
reads

\begin{widetext}
\begin{eqnarray}\label{eq:16}
{\cal L } &=& \bar{\psi}[i\gamma_{\mu}\partial^{\mu}-(M-
g_{\sigma}\phi -g_{\delta}\vec{\tau}\cdot\vec{\delta})
-g{_\omega}\gamma_\mu\omega^{\mu}-g_\rho\gamma^{\mu}\vec\tau\cdot
\vec{b}_{\mu}]\psi \nonumber \\&&
+\frac{1}{2}(\partial_{\mu}\phi\partial^{\mu}\phi-m_{\sigma}^2\phi^2)
+\frac{1}{2}m^2_{\omega}\omega_{\mu} \omega^{\mu}
+\frac{1}{2}m^2_{\rho}\vec{b}_{\mu}\cdot\vec{b}^{\mu} \nonumber
\\&&
+\frac{1}{2}(\partial_{\mu}\vec{\delta}\cdot\partial^{\mu}\vec{\delta}
-m_{\delta}^2\vec{\delta^2}) -\frac{1}{4}F_{\mu\nu}F^{\mu\nu}
-\frac{1}{4}\vec{G}_{\mu\nu}\vec{G}^{\mu\nu},
\end{eqnarray}
\end{widetext}

\noindent
where $\phi$ is the $\phi$-meson field,
 $\omega_{\mu}$ the $\omega$-meson field,
 $\vec{b}_{\mu}$  the $\rho$ meson field and $\vec{\delta}$  the isovector
scalar field
 of the $\delta$-meson, respectively.
$F_{\mu\nu}\equiv\partial_{\mu}\omega_{\nu}-\partial_{\nu}\omega_{\mu}$
 and  $\vec{G}_{\mu\nu}\equiv\partial_{\mu}\vec{b}_{\nu}-\partial_{\nu}
\vec{b}_{\mu}$.

The most important difference to conventional RMF theory is the contribution
from the rearrangement self-energies to the DDRH baryon field equation.
The meson-nucleon couplings $g_{\sigma}$,
$g_{\omega}$, $g_{\rho}$ and $g_{\delta}$ are assumed to be vertex functions
of Lorentz-scalar bilinear forms of the nucleon field operators.
In most applications of DDRH theory, these couplings are chosen as
functions of the
vector density $\hat{\rho}^{2}=\hat{j}_{\mu}\hat{j}^{\mu}$ with
$\hat{j}_{\mu}=\bar{\psi} \gamma_{\mu}\psi$.

The variational derivative of Lagrangian density Eq. (A16) is

\begin{eqnarray}\label{eq:17}
\frac{\partial \cal L}{\partial \bar{\psi}}=\frac{\partial \cal L}{\partial
\bar{\psi}}
+\frac{\partial \cal L}{\partial \hat{\rho}}\frac{\partial \hat{\rho}}
{\partial \bar{\psi}}.
\end{eqnarray}

The density dependence of  the vertex functions $g_{\sigma}$,
$g_{\omega}$, $g_{\rho}$ and $g_{\delta}$ produces the rearrangement
contribution $\Sigma_{\mu}^{R}$

\begin{eqnarray}\label{eq:18}
\Sigma^{R}=\bar{\psi}[\frac{\partial g_\sigma}{\partial \hat{\rho}}\phi
+\frac{\partial g_\delta}{\partial \hat{\rho}}\vec{\tau}\cdot\vec{\delta}
-\frac{\partial g_\omega}{\partial \hat{\rho}}\gamma_{\mu}\omega^{\mu}
-\frac{\partial g_\rho}{\partial \hat{\rho}}\gamma^{\mu}\vec{\tau}\cdot\vec
{b_{\mu}}]\psi,
\end{eqnarray}

\noindent
and
\begin{eqnarray}\label{eq:19}
\Sigma_{\mu}^{R}=-\Sigma^{R}~u_{\mu},
\end{eqnarray}

\noindent
where $u^{\mu}$=$\frac{\hat{j}^{\mu}}{\hat{\rho}}$ is a four-velocity with
$u^{\mu}u_{\mu}$=1.
The Dirac equation takes the conventional form

\begin{eqnarray}\label{eq:20}
[\gamma_{\mu}(i\partial^{\mu}-\Sigma^{\mu})-(M-\Sigma^{s})]\psi=0,
\end{eqnarray}

\noindent
where $\Sigma_{\mu}=g_{\omega}\omega_{\mu}+g_{\rho}\vec{\tau}\cdot\vec{b}_{\mu}
-\Sigma_{\mu}^{R}$ and $\Sigma_{s}=g_{\sigma}\phi+g_{\delta}
\vec{\tau}\cdot\vec{\delta}$.
The field equation of baryon in the mean-field approximation (MFA) is

\begin{eqnarray}\label{eq:21}
&& (i\gamma_{\mu}\partial^{\mu}-(M- g_{\sigma}\phi
-g_\delta{\tau_3}\delta_3)-g_\omega\gamma^{0}{\omega_0}-g_\rho\gamma^{0}
{\tau_3}{b_0}
+\gamma^0\Sigma_0^R)\psi=0,\nonumber \\
\end{eqnarray}

\noindent
with

\begin{eqnarray}\label{eq:22}
&& \phi=\frac{g_\sigma}{m_{\sigma}^2} \rho_{s}
\equiv \frac{g_\sigma}{m_{\sigma}^2} (\rho_{sp}+\rho_{sn}), \nonumber \\
&& \omega_{0}=\frac{g_\omega}{m^2_{\omega}} <\bar\psi{\gamma^0}\psi>
=\frac{g_\omega}{m_{\omega}^2}\rho
\equiv \frac{g_\omega} {m_{\omega}^2}(\rho_p+\rho_n),\nonumber \\
&& b_{0}=\frac{g_\rho}{m^2_{\rho}} <\bar\psi{\gamma^0}\tau_3\psi>
=\frac{g_\rho}{m^2_{\rho}}\rho_3
\equiv \frac{g_\rho}{m^2_{\rho}}(\rho_p-\rho_n),\nonumber \\
&& \delta_3=\frac{g_\delta}{m^2_{\delta}}<\bar\psi\tau_3\psi>
=\frac{g_{\delta}}{m^2_{\delta}} \rho_{s3}
\equiv \frac{g_{\delta}}{m^2_{\delta}}(\rho_{sp}-\rho_{sn}),\nonumber \\
&&\Sigma_0^{R}=(\frac{\partial g_\sigma}{\partial \rho})\frac{g_\sigma}
{m_\sigma^2}\rho_s^2
+(\frac{\partial g_\delta}{\partial \rho}) \frac{g_\delta}{m_\delta^2}
\rho_{s3}^2
-(\frac{\partial g_\omega}{\partial \rho}) \frac{g_{\omega}}{m_{\omega}^2}
 \rho^2
-(\frac{\partial g_\rho}{\partial \rho}) \frac{g_{\rho}}{m_{\rho}^2}
\rho_{3}^{2}.
\end{eqnarray}

where $\rho_i (i=p, n)$ and $\rho_{si}$ are  baryon and scalar nucleon 
densities, respectively.

The EoS at finite temperature in the MFA is given by

\begin{equation}\label{eq:23}
\epsilon=\sum_{i=n,p}{2}\int \frac{{\rm d}^3k}{(2\pi)^3}E_{i}^\star(k)
(f_i(k)+\bar{f}_{i}(k))
+\frac{1}{2}\frac{g_\sigma^2}{m_\sigma^2}\rho_s^2
+\frac{1}{2} \frac{g_{\omega}^2}{m_\omega^2}\rho^2
+\frac{1}{2}\frac{g_{\rho}^2}{m_{\rho}^2}\rho_3^2
+ \frac{1}{2}\frac{g_{\delta}^2}{m_{\delta}^2}\rho_{s3}^2,
\end{equation}

\begin{equation}\label{eq:24}
p =\sum_{i=n,p} \frac{2}{3}\int \frac{{\rm d}^3k}{(2\pi)^3}\frac{k^2}{E_{i}^
\star(k)}(f_i(k)+\bar{f}_{i}(k))
-\frac{1}{2}\frac{g_\sigma^2}{m_\sigma^2}\rho_s^2
+\frac{1}{2}\frac{g_{\omega}^2}{m_\omega^2}\rho^2
+\frac{1}{2}\frac{g_{\rho}^2}{m_{\rho}^2}\rho_3^2
-\frac{1}{2}\frac{g_{\delta}^2}{m_{\delta}^2}\rho_{s3}^2-\Sigma_{0}^{R}\rho.
\end{equation}

\noindent
where ${E_i}^\star=\sqrt{k^2+{{M_i}^\star}^2}$, i=p,n.

The proton and neutron chemical potentials in DDRH model are
\noindent

\begin{eqnarray}\label{eq:25}
\mu_p=\mu_p^\star + \frac{g_\omega^2}{m_\omega^2}\rho + \frac{g_\rho^2}
{m_\rho^2}\rho_3-\Sigma_0^{R} \nonumber \\
\mu_n=\mu_n^\star + \frac{g_\omega^2}{m_\omega^2}\rho - \frac{g_\rho^2}
{m_\rho^2}\rho_3-\Sigma_0^{R}.
\end{eqnarray}

and the baryon and isospin chemical potentials in the hadron sector
consistently defined like before, Eq.(\ref{eq.14}).

\subsubsection*{A2.1 Parameters of the meson-nucleon coupling density 
dependence}

The parameters of the model include  nucleon mass $M=939 MeV$, the masses of
the mesons
$m_{\sigma}$, $m_{\omega}$, $m_{\rho}$, $m_{\delta}$ and the density dependent
 meson-nucleon couplings.
The parametrization was proposed \cite{TW99} as :

\begin{equation}\label{eq:26}
g_{i}(\rho)=g_{i}(\rho_0)f_{i}(x), ~~~~for ~~i=\sigma,\omega,\rho,\delta,
\end{equation}

\noindent
with

\begin{equation}\label{eq:27}
f_{i}(x)=a_{i}\frac{1+b_{i}(x+d_i)^2}{1+c_{i}(x+d_i)^2},~~~~ for~~i=\sigma,
\omega,
\end{equation}

\noindent
where $x=\rho/\rho_0$ and $\rho_0$ is the saturation density.
For $\rho$ and $\delta$ mesons, the following parametrization was proposed
\cite{avan04} :

\begin{equation}\label{eq:28}
f_{i}(x)=a_{i}exp[-b_{i}(x-1)] - c_{i}(x-d_i), ~~~~ for~~ i=\rho,\delta.
\end{equation}

\noindent
The parametrization form and parameters are taken from ref. \cite{TW99}
for $\sigma$, $\omega$ mesons and ref. \cite{avan04} for $\rho$, $\delta$
mesons, respectively.
 All parameters  are listed in Table 3.

\par
\noindent

\begin{center}
{{\large \bf Table 3.}~~Parameters of the DDRH model}.
\par
\vspace{0.5cm} \noindent

\begin{tabular}{c|c|c|c|c|c} \hline
 Model  &\multicolumn{2}{c|} {$TW~[17]$} &DDRH$\rho$~[28]   &\multicolumn{2}{c}
{DDRH$\rho\delta$~[28]} \\ \hline
 Meson          &$\sigma$   &$\omega$   &$\rho$           &$\rho$    &$\delta$
\\ \hline
$m_i~(MeV)$     &550        &783        &770              &770       &980
\\ \hline
$g_i(\rho_{0})$ &10.72854   &13.29015   &3.587            &6.476     &7.58963
\\ \hline
$a_i$           &1.365469   &1.402488   &0.095268         &0.095268  &0.01984
\\ \hline
$b_i$           &0.226061   &0.172577   &2.171            &2.171     &3.4732
\\ \hline
$c_i$           &0.409704   &0.344293   &0.05336          &0.05336   &-0.0908
\\ \hline
$d_i$           &0.901995   &0.983955   &17.8431          &17.8431   &-9.811
\\ \hline
\end{tabular}
\end{center}

\noindent
\begin{figure}[hbtp]
\begin{center}
\includegraphics[scale=0.4]{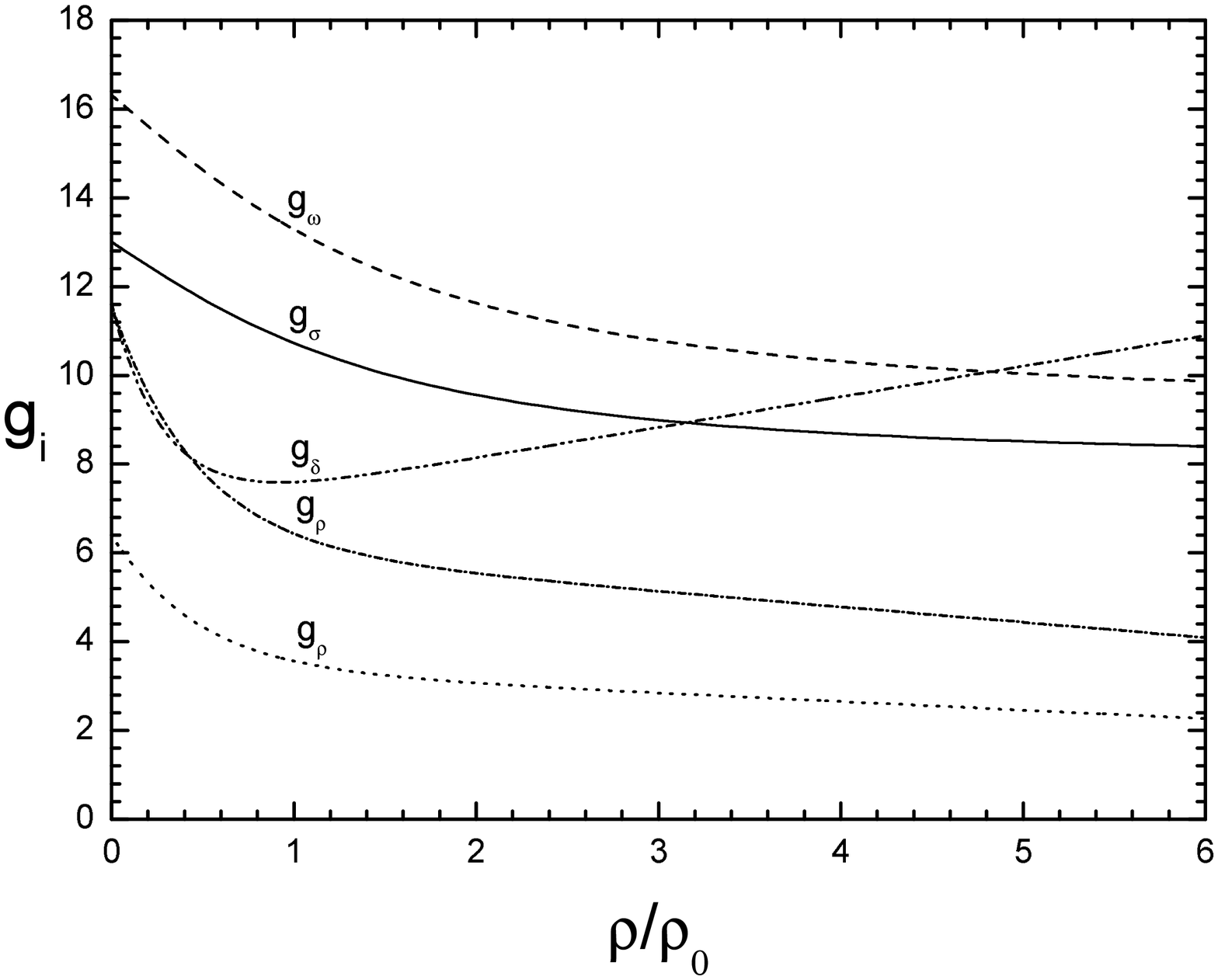}
\caption{Density dependence of the meson-nucleon couplings in the used 
DDRH interactions. The bottom $g_\rho$ dashed line corresponds to the 
$DDRH \rho$ case, without the $\delta$ meson.}
\label{couplings}
\end{center}
\end{figure}

The corresponding density dependence of the couplings is presented in the 
Fig.\ref{couplings}.

\noindent

\begin{figure}[hbtp]
\begin{center}
\vglue +3.0cm
\includegraphics[scale=0.4]{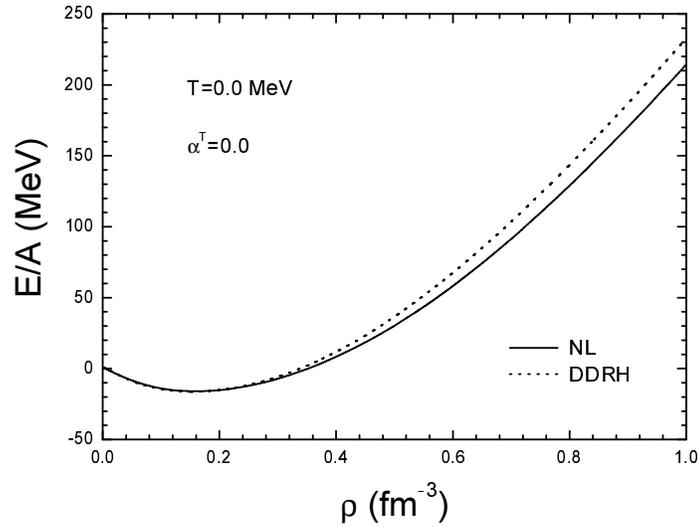}
\vglue -4.0cm
\caption{Energy per nucleon as a function of the baryon density
for symmetric matter at T=0. Solid line: NL results. Dashed line: DDRH.}
\label{eacomp}
\end{center}
\end{figure}

\noindent

\begin{figure}[hbtp]
\begin{center}
\includegraphics[scale=0.5]{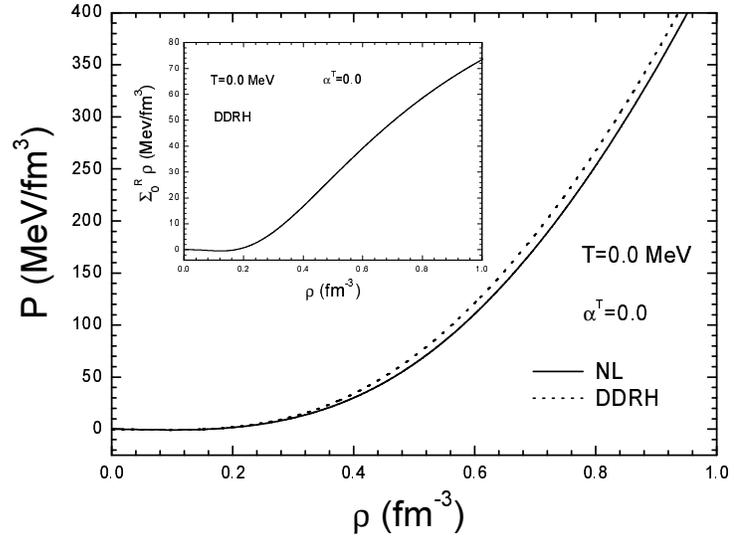}
\vglue -5.0cm
\caption{Pressure as a function of the baryon density
for symmetric matter at T=0. Solid line: NL results. Dashed line: DDRH.
In the inset we show the density dependence of the rearrangement term
correction to the DDRH pressure}
\label{psigma}
\end{center}
\end{figure}

\begin{figure}[hbtp]
\begin{center}
\vglue +3.0cm
\includegraphics[scale=0.4]{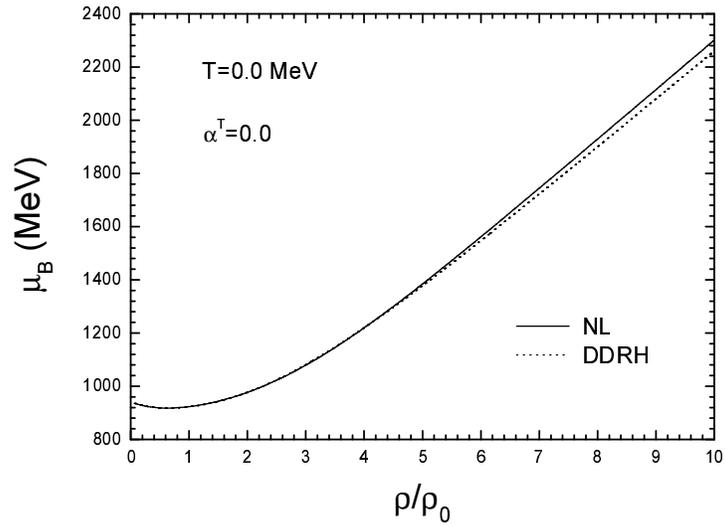}
\vglue -3.0cm
\caption{Baryon chemical potential as a function of the baryon density
for symmetric matter at T=0. Solid line: NL results. Dashed line: DDRH.}
\label{mubrho}
\end{center}
\end{figure}

\begin{figure}[hbtp]
\begin{center}
\includegraphics[scale=0.4]{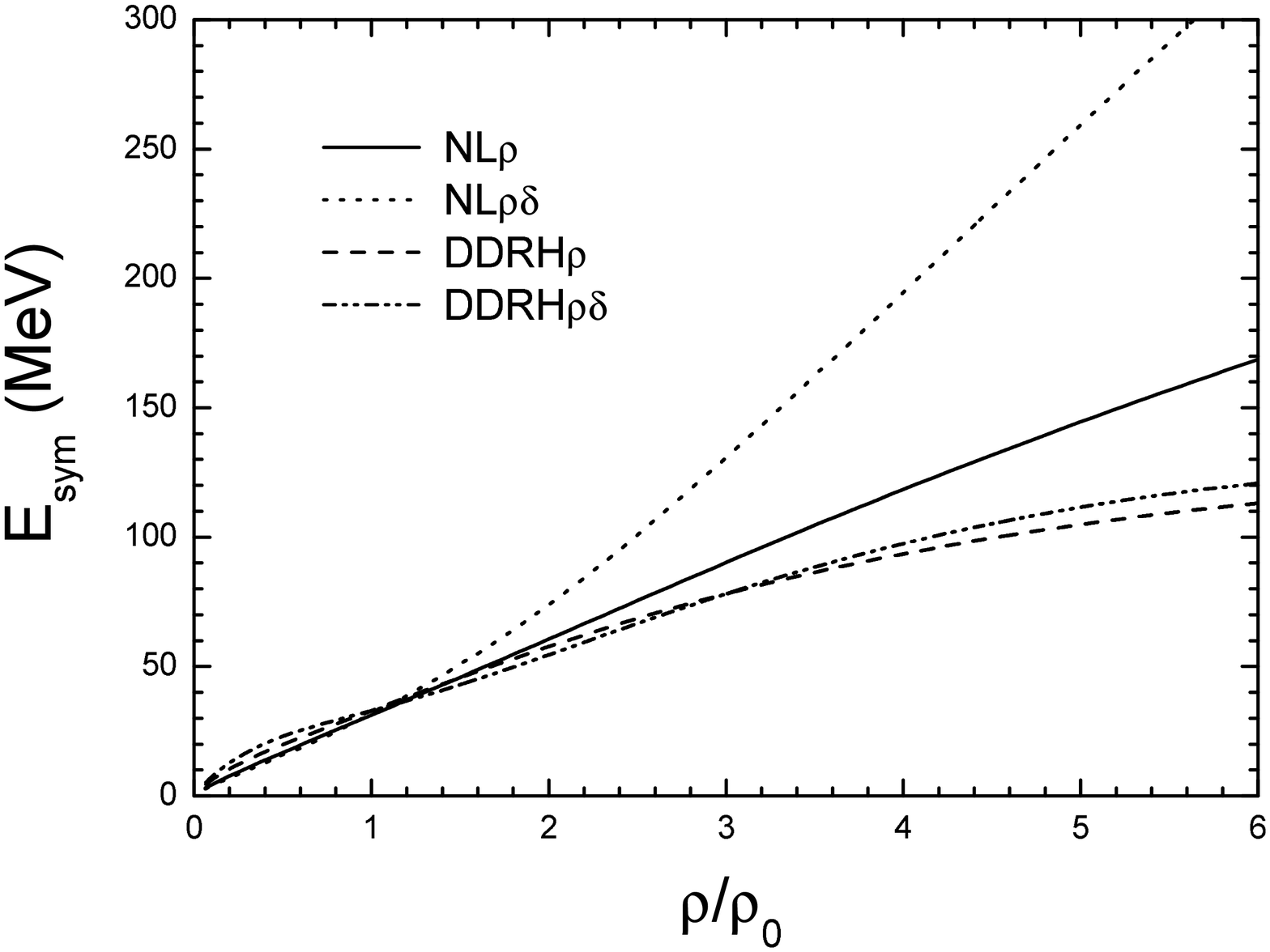}
\caption{Symmetry energy as a function of the baryon density at T=0. Solid 
and dotted lines: $NL\rho-\rho\delta$ results with constant couplings. 
Dashed and dot-dashes lines: $DDRH$ results with density dependent couplings.}
\label{esym}
\end{center}
\end{figure}


\subsubsection*{A2.2 Hadronic EoS without and with density dependent couplings}

In order to better understand the effects of the density dependence of the
nucleon-meson couplings on the hadron-quark phase transition we look here in
more detail at the differences on the hadron EoS, $NL$ vs $DDRH$ approaches.
In Figs. \ref{eacomp}, \ref{psigma}, and \ref{mubrho}, we show 
in the order the results for the energy per
particle, the pressure and the baryon chemical potential 
in the case of isospin symmetric case at zero temperature.
With density dependent couplings we
see a more repulsive hadronic matter and so in general we would expect
an ``earlier''
transition to the quark phase with increasing density.

In any case for symmetric matter the effect will not be too large. In fact we
note that the DDRH pressure
is not much larger that the NL values. The same is happening for the
baryon chemical potentials. This is mostly due to the increase
with density of the rearrangement corrections $\Sigma_0^R \rho$,
Eq.(\ref{eq:24}),
as shown in the inset of  Fig.\ref{psigma}.

At variance the effect of density dependent couplings  will be 
noticeably larger 
in the case of isospin asymmetric
matter. In fact at high baryon density the symmetry term will be much
reduced due to
the decrease of the $\rho-nucleon$ coupling, see Fig.\ref{couplings}.
This can be clearly observed from the Fig.\ref{esym}.

\section{Quark matter equation of state}
{\hskip  0.7cm}

The energy density and the pressure for the quark system are,
respectively, given by the MIT Bag model \cite{MIT}


\begin{eqnarray}\label{eq:29}
\epsilon= 3 \times 2 \sum_{q=u,d,s}\int \frac{{\rm d}^3k}
{(2\pi)^3}\sqrt{k^{2}+m_{q}^{2}}(f_{q}+\bar{f}_{q})
+B~,
\end{eqnarray}

\begin{eqnarray}\label{eq:30}
 P =\frac{ 3\times 2}{3}\sum_{q=u,d,s}\int \frac{{\rm d}^3k}{(2\pi)^3}
\frac{k^2}{\sqrt{k^{2}+m_{q}^{2}}} (f_{q}+\bar{f}_{q})
-B~,
\end{eqnarray}

\noindent
where B denotes the Bag constant (the Bag pressure) chosen as a rather
standard parameter \cite{s10}. In fact a detailed discussion about the effects
of different Bag constants on the hadron-quark transition will be the
main topic of this paper.
$m_{q}$, $q=u, d$, are the free quark masses ($5.5~MeV$ choice), and
$f_q$, $\bar{f}_{q}$ represent
the Fermi distribution functions for quarks and anti-quarks,
respectively :

\noindent
\begin{eqnarray}\label{eq.31}
f_q=\frac{1}{1+\exp\{({E_q}-{\mu_q})/T \}}\,,
\end{eqnarray}

and

\noindent
\begin{eqnarray}\label{eq.32}
\bar{f}_{q}=\frac{1}{1+\exp\{({E_q}+{\mu_q})/T \}}.
\end{eqnarray}

\noindent
where $E_{q}=\sqrt{k^{2}+m_{q}^{2}}$ and $\mu_{q}$ are the chemical potentials
for quarks and anti-quarks of type q.

The quark number density is given by
\noindent

\begin{eqnarray}\label{eq:33}
 n_{i}=<q_{i}^{+}q_{i}>={3\times 2} \int \frac{{\rm d}^3k}{(2\pi)^3}
 (f_{i}-\bar{f}_{i})~, ~~~~i=u,d,
\end{eqnarray}

\noindent
where the chemical potential ${\mu_{q}}$
can be obtained by Eq. (B5) for a given quark number density.
The quark number density is $n_{q}=n_{u}+n_{d}$ $\equiv$ $\rho_q$ and 
relation with the
baryon density is
\noindent

\begin{eqnarray}\label{eq:34}
\rho_B^{Q}=\frac{1}{3} \rho_q=\frac{1}{3}(\rho_{u}+\rho_{d}).
\end{eqnarray}

The chemical potentials of quarks are related to the baryon and isospin
chemical potentials

\begin{eqnarray}\label{eq:35}
\mu_u=\frac{1}{3} \mu_B + \mu_3, ~~~~~\mu_d=\frac{1}{3} \mu_B - \mu_3.
\end{eqnarray}

\vspace {1.5cm}



\begin{thebibliography}{s4}

\bibitem{Shao10}G. Y. Shao and Y. X. Liu, Phys. Rev. C {\bf 82}, 055801 (2010).

\bibitem{Xu10}J. Xu, L. W. Chen, C. M Ko, and B. A. Li, Phys. Rev. C {\bf 81},
 055803 (2010).

\bibitem{muller} H. M\"{u}ller, Nucl. Phys. A {\bf 618}, 349 (1997).

\bibitem{ditoro1} M.Di Toro, A. Drago, T. Gaitanos, V, Greco, A. Lavagno,
                  Nucl. Phys. A {\bf 775}, 102 (2006)

\bibitem{ditoro2} M.Di Toro, B. Liu, V. Greco, V. Baran, M. Colonna, and
S. Plumari, Phys. Rev. C {\bf 83},  014911 (2011).

\bibitem{Pagliara10}G. Pagliara and J. Schaffner-Bielich, Phys. Rev. D
{\bf 81}, 094024 (2010).

\bibitem{Cavagnoli10}R. Cavagnoli, C. Provid\^{e}ncia, and D. P. Menezes,
arXiv:1009.3596v1.

\bibitem{SW85} B.D. Serot and J.D. Walecka, Adv. Nucl. Phys. 
{\bf 16}, 1 (1985).

\bibitem{bog97} J. Boguta and A.R. Bodmer, Nucl. Phys. A {\bf 292}, 413 (1997).


\bibitem{liubo02} B. Liu, V. Greco, V. Baran, M. Colonna, and M. Di Toro,\\
             Phys. Rev. C {\bf 65}, 045201 (2002).

\bibitem{menpro04} D.P.Menezes and C. Provid\^{e}ncia,
             Phys. Rev. C {\bf 70}, 058801 (2004).

\bibitem{baranPR} V. Baran, M. Colonna, V. Greco, M. Di Toro,
             Phys. Rep. {\bf 410}, 335 (2005).

\bibitem{gait04} T. Gaitanos, M. Di Toro, S. Typel, V. Baran, C. Fuchs,
                 V. Greco, H.H. Wolter, Nucl. Phys. 
                  A {\bf 732}, 24 (2004).

\bibitem{liubo05} B. Liu, H. Guo, M. Di Toro, and V. Greco,
                  Eur. Phys. J. A {\bf 25}, 293 (2005).

\bibitem{fuchs95} C.Fuchs, H. Lenske, and H.H. Wolter,
                  Phys. Rev. C {\bf 52}, 3043 (1995).

\bibitem{jong98}F. de Jong, H. Lenske, Phys. Rev. C57, 3099 (1998); 
 F. Hofmann, C.M. Keil, H. Lenske, Phys. Rev. C {\bf 64}, 034314 (2001).


\bibitem{TW99} S.Typel and H.H. Wolter,
               Nucl. Phys. A {\bf 656}, 331 (1999).

\bibitem{vandal04}E.N.E. van Dalen, C. Fuchs, Amand Faessler,
                 Nucl. Phys. A {\bf 744}, (2004) 227; 
                 Phys.Rev. C {\bf 72}, (2005) 065803.

\bibitem{liubo06} B. Liu, M. Di Toro, V. Greco, C.W. Shen, E.G. Zhao and
                  B.X. Sun, Phys. Rev. C {\bf 75}, 048801 (2007).

\bibitem{shao11} G.Y. Shao, M. Di Toro, B. Liu, M. Colonna, V. Greco, Y.X. Liu,
 \emph{Hadron-quark transition in asymmetric matter with dynamical
quark masses}, arXiv:1102.4964v1[nucl-th].

\bibitem{erice08} 
M. Di Toro et al. Progr.Part.Nucl.Phys. {\bf 62}, 389-401 (2009).


\bibitem{greco03}
V. Greco et al., Phys.Lett. {\bf B562}, 215 (2003).

\bibitem{ferini06}
G. Ferini, T. Gaitanos, M. Colonna, M. Di Toro. H.H. Wolter,
Phys.Rev.Lett. {\bf 97}, 202301 (2006).

\bibitem{baoPR}
B.A. Li, L.W. Chen, C.M. Ko, Phys.Rep. {\bf 465}, 113 (2008).

\bibitem{Bmuller85}
B. M\"{u}ller,
 \emph{The Physics of the Quark-Gluon-Plasma}, Lecture Notes in Physics,
 Vol. 225, Springer-Verlag, Berlin-Heidelberg 1985.

\bibitem{Yagi05}
K. Yagi, T. Hatsuda, Y. Miaka,
\emph{Quark Gluon Plasma}, Ch.3, Cambridge Unversity Press 2005.

\bibitem{venu92}
R. Venugopalan, M. Prakash,
Nucl. Phys. A {\bf 546}, 718 (1992).

\bibitem{bora10}
S.Boranyi, Z. Fodor, C. Hoelbling, S.D. Katz, S. Krieg, C. Ratti, K.K. Szabo,
 JHEP {\bf 09}, 073 (2010); arXiv:1005.3508.

\bibitem{pressure}
We remark that the opposite is seen in the $P~-~\rho_B$ plane. In fact
repulsive vector interaction terms are increasing the pressure for a fixed 
density but meanwhile also the chemical potential is increasing,
 see the Appendix A.

\bibitem{greco}V. Greco, M. Colonna, M.Di Toro, F. Matera,
            Phys. Rev. C {\bf 67}, 015203 (2003).

\bibitem{avan04}S.S Avancini, L. Brito, D.P. Menezes, and C. Providencia,
                Phys. Rev. C {\bf 70}, 015203 (2004).

\bibitem{MIT}   A. Chodos, et al., Phys, Rev. D {\bf 9}, 3471 (1974).

\bibitem{s10} K. Schertler, C. Greiner, P.K. Sahu, M.H. Thoma,
          Nucl. Phys. A {\bf 637}, 451 (1998).

\end{thebibliography}
\end{document}